\documentstyle[sprocl,epsf,fleqn]{article}

\def\Journal#1#2#3#4{{#1} {\bf #2}, #3 (#4)}

\def\PLB{{\em Phys. Lett.}  B}

\def\PRD{{\em Phys. Rev.} D}

\def\ra{\rightarrow}

\def\be{\begin{equation}}
\def\ee{\end{equation}}
\def\bea{\begin{eqnarray}}
\def\eea{\end{eqnarray}}

\newcommand{\AmS}{{\protect\the\textfont2
  A\kern-.1667em\lower.5ex\hbox{M}\kern-.125emS}}

\def \Rpv{R_{P} \hspace{-1.2em}/\;\hspace{0.2em}}

\begin{document}

\title{NEW PHYSICS POTENTIAL OF DOUBLE BETA DECAY AND DARK MATTER
SEARCH
\footnote{Talk presented by Heinrich P\"as at the
at the 6th Symp. on Particles, Strings and Cosmology
(PASCOS'98), Boston, March 1998,\\ 
E-mail: Heinrich.Paes@mpi-hd.mpg.de; 
Web:http://pluto.mpi-hd.mpg.de/$~$paes/heini.html}}

\author{H.V. Klapdor--Kleingrothaus, H. P\"as}

\address{Max--Planck--Institut f\"ur Kernphysik\\ 
P.O. Box 103980, D--69029 Heidelberg, Germany} 

\maketitle\abstracts{ 
The search for
neutrinoless double beta decay and WIMP dark matter
has a broad potential 
to test particle physics beyond the standard model. 
During the last years, the analysis of various contributions to the
double beta decay rate
by the Heidelberg group led, 
besides the most restrictive limit on the effective 
Majorana neutrino mass,
to bounds on 
left-right-symmetric models, leptoquarks and supersymmetry.
In a general framework bounds on arbitrary lepton number violating theories 
can be derived.
Using double beta technology for direct dark matter detection, 
stringent limits on the spin-independent WIMP--nucleon interaction
have been obtained.  
These results deduced from the Heidelberg-Moscow double 
beta decay experiment are reviewed. Also an outlook on the future of double 
beta decay and dark matter search, the GENIUS proposal, is given.
}

\section{Double Beta Decay -- the basic mechanism}
Double beta decay ($0\nu\beta\beta$)
\cite{Kla98,tren} corresponds to two single beta decays 
occuring in 
one
nucleus and 
converts a nucleus (Z,A) into a nucleus (Z+2,A).
While even the standard model (SM) allowed process emitting two antineutrinos
\be
^{A}_{Z}X \rightarrow ^A_{Z+2}X + 2 e^- + 2 {\overline \nu_e}
\ee
is one of the rarest processes in nature with half lives in the region of
$10^{21-24}$ years, more interesting is the search for 
the neutrinoless mode,
\be        
^{A}_{Z}X \rightarrow ^A_{Z+2}X + 2 e^- 
\ee
which
violates lepton number by two units and thus implies physics beyond the 
SM.

\section{The Heidelberg--Moscow Double Beta Decay Experiment}
The Heidelberg--Moscow experiment \cite{HM97,HM97b,Kla98}
is a second generation experiment
searching for the $0\nu\beta\beta$ decay of $^{76}$Ge. 
Five crystals with an active mass of 10.96 kg, 
grown out of 19.2 kg of 86\% enriched $^{76}$Ge, are in regular operation
as p--type HPGe detectors in low level cryostats in the Gran Sasso laboratory,
which provides a shielding of 3500 m of water equivalent (mwe).
The high source strength of the experiment, the large size of the detectors 
concentrating the
background in the peaks and the excellent energy resolution yield an
outstanding position compared with other experiments (see Fig. 1).
It has been described recently in detail in \cite{Kla98,HM97,HM97b}.

Fig. 2 shows the results after 35 kg y
measuring time for all data corresponding to a half life limit of
\be
T_{1/2}^{0\nu\beta\beta}>1.2 \cdot 10^{25} y.
\ee 
The limits from the pulse shape 
\cite{hel,HM97b}
data with 18 kg y (filled histogram in Fig. 
2), are just becoming competitive 
to the large data set without application of pulse shape analysis.
The background improvement will allow to test the half life 
region up to $6 \cdot 10^{25}$ y, corresponding to a neutrino mass limit of 
0.1--0.2 eV, during the next five years.

\begin{figure}[!t] 
\vspace*{-20mm}
\hspace*{1cm}
\epsfxsize=80mm
\epsfbox{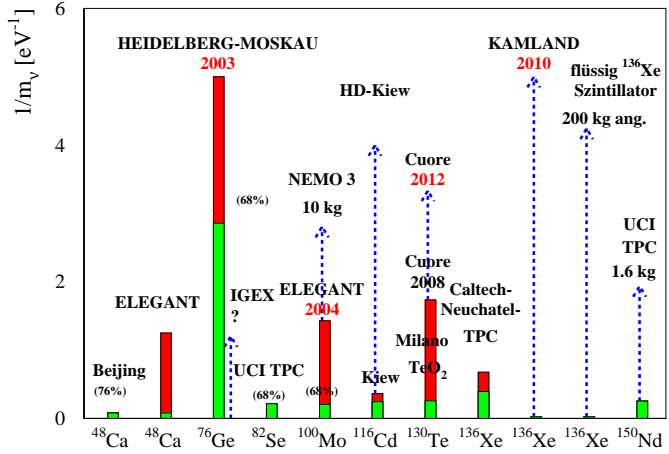}
\vspace*{-30mm}
\caption{Comparison of running and proposed double beta  experiments.
Shown is the sensitivity on the effective neutrino Majorana mass.
The filled bars correspond to the present status, open bars correspond
to `safe' expectations for the year 2003 and dashed lines corresond to 
long--term planned or hypothetical experiments.}
\vspace*{-5mm}
\end{figure}

The SM allowed $2\nu\beta\beta$ decay is measured with  
high statistics, containing 21115 events in the energy region
of 500-2040 keV, and yields a half life of \cite{HM97}
\be
T_{1/2}^{2\nu\beta\beta}=(1.77^{+0.01}_{-0.01}(stat.)^{+0.13}_{-0.11}(syst.)) 
\cdot 10^{21} y.
\ee
This result, confirming the theoretical predictions of 
\cite{staudt}
with an accuracy of
a factor $\sim \sqrt{2}$, provides a consistency check 
of nuclear matrix 
element calculations. 
It also for the first time opens up the possibility to 
search for deviations of the $2\nu\beta\beta$ spectrum such as those due to
emission of exotic scalars \cite{major}.

\begin{figure}[!t]
\hspace*{3mm}
\epsfxsize=80mm
\epsfbox{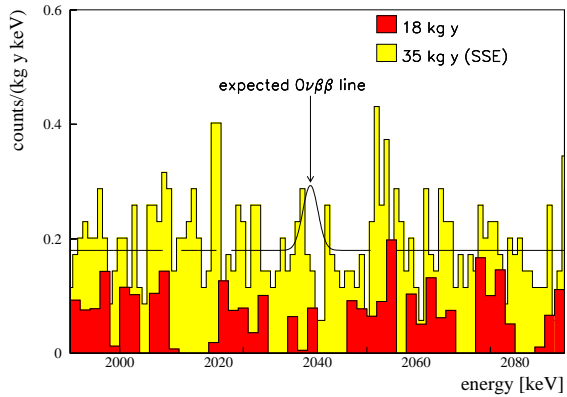}
\caption{Combined spectrum in the $0\nu\beta\beta$ peak region: 
Complete data (35 kg $\cdot$y),
with pulse shape analysis detected single site events (18 kg 
$\cdot$ y) and
90\% C.L. excluded peak.}
\vspace*{-5mm}
\end{figure}

\section{Double Beta Decay and Physics Beyond the Standard Model}

\subsection{Neutrino Mass}
The search for $0\nu\beta\beta$ decay exchanging a massive 
left--handed Majorana
neutrino between two SM vertices 
at present provides the most 
sensitive approach to determine an absolute neutrino mass and also 
a unique possibility to distinguish between the Dirac or  
Majorana nature of the neutrino.
With the recent half life limit of the Heidelberg--Moscow
experiment \cite{HM97b,Kla98} the following limits on effective left--handed 
neutrino masses can be deduced:
\be
\langle m_{\nu}\rangle \leq 0.44 eV \hskip5mm (90\% C.L.)
\ee  
\be
\langle m_{\nu}\rangle \geq 7.5\cdot 10^7 GeV \hskip5mm(90\% C.L.)
\ee
Taking into account the uncertainties in the numerical values of
nuclear matrix elements of about a
factor of 2, the Heidelberg--Moscow experiment, improving its half life limit
up to $6 \cdot 10^{25}y$, will test degenerate neutrino 
scenarios \cite{deg} in the next five years. 

\subsection{Left--Right--Symmetric Models}
In left--right symmetric models the left--handedness of weak 
interactions is explained as due to the effect of different symmetry breaking 
scales in the left-- and in the right--handed sector. 
$0\nu\beta\beta$ decay proceeds through exchange of the heavy right--handed 
partner of the ordinary neutrino between right-handed W vertices, leading
to a limit of
\be
m_{WR}\geq 1.2 \Big(\frac{m_N}{1TeV}\Big)^{-(1/4)} TeV.
\ee 
Including a theoretical limit obtained from considerations of vacuum 
stability \cite{moha86} one can deduce an absolute lower limit on the 
right--handed W mass of \cite{11} 
\be
m_{W_{R}}\geq 1.2 TeV.
\ee

\subsection{Supersymmetry}
Supersymmetry (SUSY), providing a symmetry between fermions and bosons and
thus doubling the particle spectrum of the SM, belongs
to the most prominent extensions of the SM. 
While in the minimal supersymmetric extension (MSSM) R--parity
is assumed to be conserved, there are no theoretical reasons for $R_p$ 
conservation and
several GUT \cite{rpguts} and Superstring \cite{rpss} models require
R--parity violation in the low energy regime.  
Also the reports concerning an anomaly at HERA  
\cite{HERA} have renewed the interest in $\Rpv$--SUSY
(see for example \cite{kalino,drei}).
In this case $0\nu\beta\beta$ decay can occur through Feynman graphs 
involving the exchange of
superpartners as well as $\Rpv$--couplings $\lambda^{'}$ 
\cite{hir95,hir95d,hir96c,hir96,Paes98}.
The half--life limit of the Heidelberg--Moscow experiment leads to bounds
in a multidimensional parameter space \cite{hir95,hir96c}
\be
\lambda_{111}^{'}\leq 3.2\times 10^{-4}\Big(\frac{m_{\tilde{q}}}{100 
GeV} \Big)^2
\Big(\frac{m_{\tilde{g}}}{100 GeV} \Big)^{1/2}
\ee
(for $m_{\tilde{d}_{R}}=m_{\tilde{u}_{L}}$), which are the sharpest limits on
 $\Rpv$--SUSY (see Fig. 3).

In addition $0\nu\beta\beta$ decay is not only sensitive to 
$\lambda_{111}^{'}$. Taking into account the fact that the SUSY partners of the
left and right--handed quark states can mix with each other, new diagrams 
appear in which the neutrino mediated double beta decay is accompanied by
SUSY exchange in the vertices \cite{babu95,hir96}.
A calculation of previously neglected tensor contributions to the decay rate
allows to derive improved limits on different combinations of $\lambda^{'}$
\cite{Paes98}. 
Assuming the supersymmetric mass parameters of order 100 GeV, the half life 
limit of the Heidelberg--Moscow Experiment implies:
$\lambda_{113}^{'} \lambda_{131}^{'}\leq 5.8 \cdot 10^{-8}$,
$\lambda_{112}^{'} \lambda_{121}^{'}\leq  1.7 \cdot 10^{-6}$

In the case of R--parity conserving SUSY, based on a theorem proven in 
\cite{sneut},
the $0\nu\beta\beta$ mass limits can be converted in sneutrino Majorana mass
term limits being more restrictive than what could be obtained
in inverse neutrinoless double beta 
decay and single sneutrino production at future linear colliders (NLC)
\cite{sneut}.
 
\begin{figure}[t]
\vskip-25mm 
\hskip10mm
\epsfxsize=100mm
\epsfysize=120mm
\epsfbox{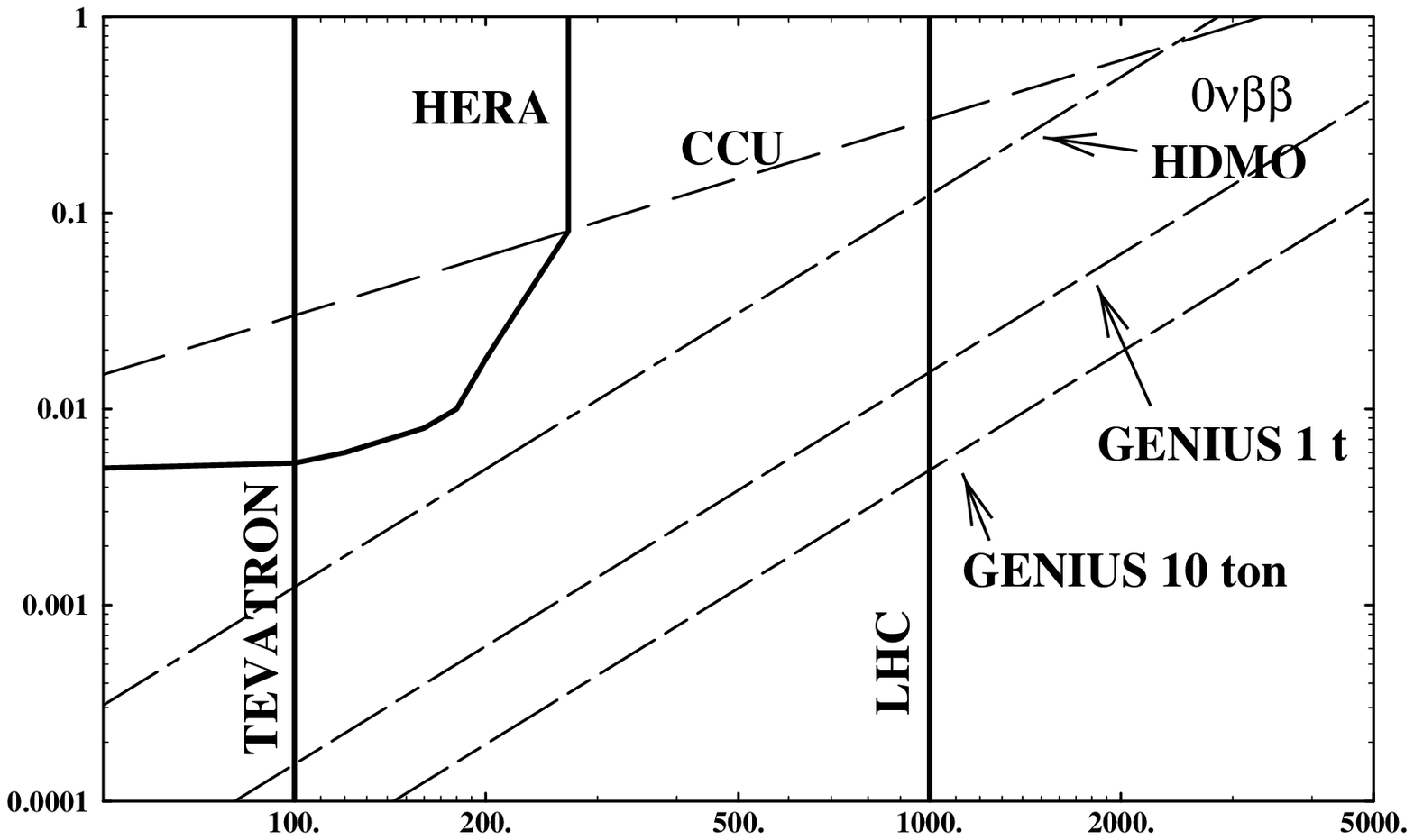}

\vskip-80mm
\noindent
$\lambda'_{111}$ 

\vskip45mm 
\hskip90mm $m_{\tilde q}$ [GeV] 

\caption{ Comparison of sensitivities of existing and future 
experiments on $\Rpv$ SUSY models in the plane $\lambda'_{111}-m_{\tilde q}$ 
(double logarithmic scale). Shown are the areas currently excluded 
by the experiments at the TEVATRON, the limit from charged-current 
universality (CCU), and the limit from absence of $0\nu\beta\beta$ 
decay from the Heidelberg-Moscow collaboration (HDMO). 
In addition, the estimated sensitivity of HERA and the LHC is compared to the 
one expected for GENIUS in the 1 ton and the 10 ton version.
(from [34])} 
\end{figure}

\subsection{Leptoquarks}
Leptoquarks are scalar or vector particles coupling both to leptons 
and quarks,
which appear naturally in GUT, extended Technicolor
or Compositeness models containing leptons and 
quarks in the same multiplet. 
Also the production  of a scalar leptoquark with mass of $m_{LQ}\simeq 200GeV$ 
has been discussed as possible effect in recent accelerator experiments
such as HERA (see for example \cite{kalino}). 
To keep the leptoquark option interesting 
in view of stringent TEVATRON limits  
($m_{LQ}>240GeV$ for scalar leptoquarks decaying with branching ratio 1
into electrons and quarks \cite{cdf}) possibilities to
reduce the branching ratio due
to the mixing of different multiplets \cite{babu97}
have been examined. This kind of mixing can be obtained by 
introducing a leptoquark--Higgs coupling -- which would lead  to a contribution
to $0\nu\beta\beta$ decay
\cite{hir96a}. Combined with the half--life limit of the 
Heidelberg--Moscow experiment bounds on effective couplings can be derived
\cite{lepbb}. 
Assuming 
only one lepton number violating $\Delta L=2$ LQ--Higgs coupling unequal 
to zero and  the leptoquark masses 
not too different, one can derive from this limit either a bound on the 
LQ--Higgs coupling
\bea
Y_{LQ-Higgs}=(few)\cdot 10^{-6}
\eea
or a limit excluding Leptoquarks with masses in the 
range of
${\cal O}(200 GeV)$. This excludes 
most of the possibilities to relax the TEVATRON bounds by introducing 
LQ--Higgs couplings to reduce the branching ratio \cite{Hir97b}.

\subsection{A general context for the double beta decay rate}
The variety of non--SM couplings appearing in different 
contributions to $0\nu\beta\beta$ decay led to the idea to construct
the general double beta decay rate allowed by Lorentz--invariance
\cite{Paes97,Paes98a}.
This approach allows to constrain lepton number violating parameters in
arbitrary models. 

For the long range part of the decay rate with two separable vertices
and light neutrino exchange in between, 
one has to consider the Lorentz-invariant contractions of six projections
with defined helicity both for the leptonic and hadronic current.
The general Lagrangian can be written in terms of
effective couplings $\epsilon^{\alpha}_{\beta}$, which correspond to the
pointlike vertices at the Fermi scale so that Fierz rearrangement is 
applicable:
\be
{\cal L}=\frac{G_F}{\sqrt{2}}\{
j_{V-A}^{\mu}J^{\dagger}_{V-A,\mu}+ \sum_{\alpha,\beta} 
 ^{'}\epsilon_{\alpha}^{\beta}j_{\beta} J^{\dagger}_{\alpha}\}
\ee
with the combinations of hadronic and leptonic Lorentz currents of 
defined helicity 
$\alpha,\beta=V-A,V+A,S-P,S+P,T_L,T_R$
The prime indicates 
the sum runs over all contractions allowed by 
Lorentz--invariance,
except for $\alpha=\beta=V-A$.
Here $\epsilon_{\alpha}^{\beta}$ denotes the 
strength of the non--SM couplings. 
For the helicity suppressed terms proportional to the (from below)
unconstrained neutrino mass no limit can be derived and terms proportional
$(\epsilon_{\alpha}^{\beta})^2$ can be neglected. 
The limits on the remaining non--SM couplings derived in in s-wave 
approximation and evaluated ``on axis'' are \cite{Paes97,Paes98a}:
$\epsilon^{V+A}_{V+A}< 7.0 \cdot 10^{-7}$, 
$\epsilon^{V+A}_{V-A}< 4.4 \cdot 10^{-9}$,
$\epsilon^{S+P}_{S+P}<1.1 \cdot 10^{-8}$,
$\epsilon^{S+P}_{S-P}<1.1 \cdot 10^{-8}$,
$\epsilon^{T_{R}}_{T_{R}}< 2 \cdot 10^{-9}$,
$\epsilon^{T_{R}}_{T_{L}}< 7\cdot 10^{-10}$.
A further step will consider the short range part of the general decay rate.

\section{WIMP Dark Matter Search with Double Beta Experiments}

Weakly interacting masssive particles (WIMPs) such as the lightest 
supersymmetric
particle (LSP) are major candidates for the cold component of nonbaryonic
dark matter in the universe. 
Due to its low background properties double beta technology can also find 
applications in the search for direct detection of WIMPs. 
The Heidelberg--Moscow Experiment, without being specially designed
for this purpose,
gave the most stringent limits on WIMPs for several years \cite{beck}. 
New results with 0.69 kg y of measurement reached a background level of 0.042 
cts/(kg d keV) in the region between 15 keV and 40 keV. The derived limit
excludes WIMPS with masses greater than 13 GeV and cross sections as low as 
$1.12 \cdot 10 ^{-5}$ pb (see Fig. 4). These are the most stringent limits on 
spin-independent interactions using only raw data \cite{HM98}. 

\begin{figure}[!t]
\hspace*{3mm}
\epsfxsize=80mm
\epsfbox{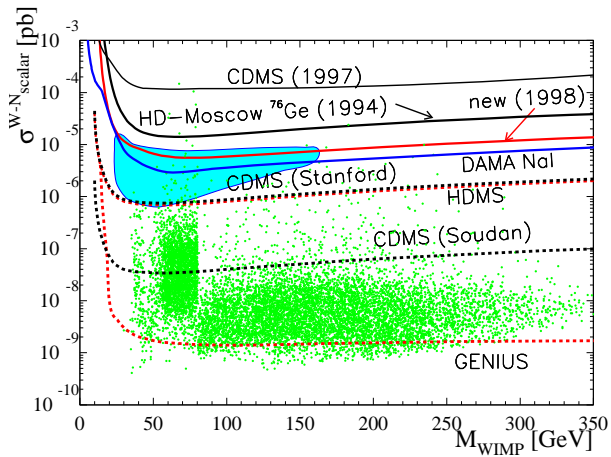 }
\caption{Compared sensitivity of running WIMP direct detection experiments and
proposals, including experiments using double beta technology (grey lines). 
Shown are limits on WIMP--nucleon 
scalar cross sections as a function of the WIMP mass. Solid lines: limits from
CDMS (1997), Heidelberg--Moscow (analysis 1994 and 1998) and DAMA
as well as their evidence contour. 
Dashed lines: Expected sensitivity of upcoming experiments, 
the Heidelberg project HDMS,
CDMS (Stanford and Soudan) as well as the Heidelberg proposal GENIUS
(100 kg version). In light grey a scatter plot calculated in the MSSM 
framework with non--universal scalar mass unification
(from [33])}.
\vspace*{-5mm}
\end{figure}

\section{Outlook on the future: GENIUS}
To render possible a further breakthrough in search for neutrino masses and 
physics beyond the SM, recently GENIUS, an experiment operating
a large amount of naked Ge--detectors in a liquid nitrogen shielding,
has been proposed \cite{Kla98}, and studied in detail in \cite{gen,genb}.
The possibility to operate Ge detectors inside liquid nitrogen has already been
demonstrated by the Heidelberg group. An excellent energy resolution 
and threshold is obtained.

Operating 288 enriched $^{76}$Ge detectors with a total mass of 1  ton 
inside a nitrogen tank of 11-12 m height and diameter, improves 
the sensitivity to neutrino masses down to 0.01 eV. This  allows to 
definitly exclude
$\nu_{e} \leftrightarrow \nu_{\mu}$ oscillations
as solution for the 
atmospheric neutrino problem (already disfavored

\newpage
\begin{figure}[!hb] 
\vspace*{-20mm}
\hspace*{-2cm}
\epsfxsize=170mm
\epsfbox{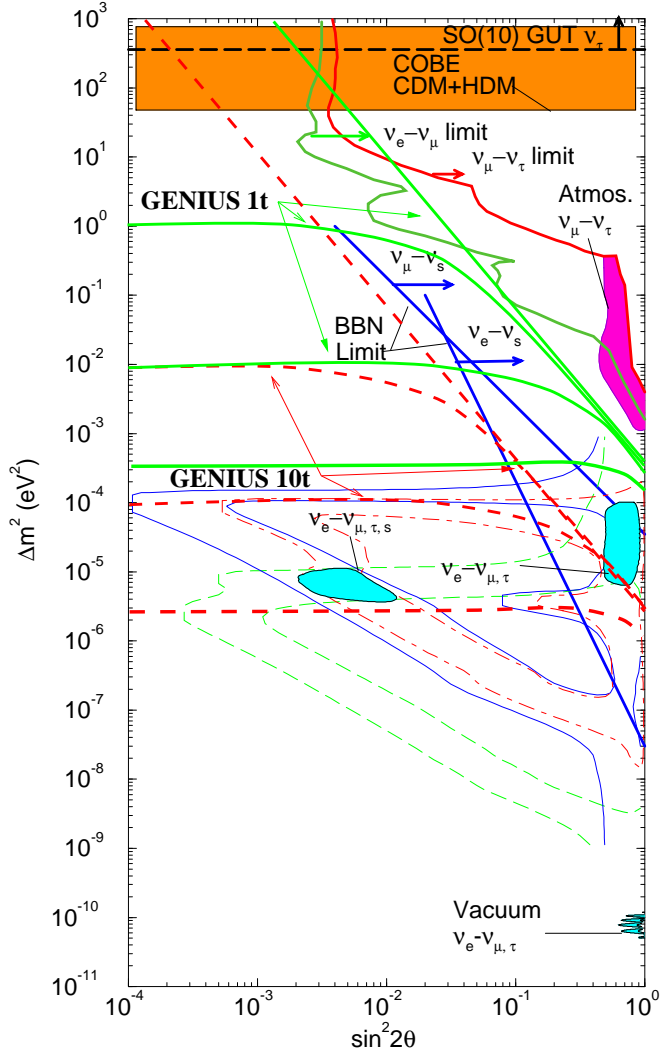}
\vspace*{-5mm}
\caption{Sensitivity of the GENIUS proposal on  the parameters $\Delta m^2$
and $\sin^2 2 \theta$, compared with various hints and bounds 
from neutrino oscilllation experiments. 
The solid lines indicate the sensitivity of the 1 ton version, 
the dashed lines for the 10 ton version, both 
for different hierarchical models with neutrino mass ratios of
(from top to bottom) 
$m_1/m_2=0,0.01,0.1,0.5$ (from [31]).} 
\vspace*{-8mm}
\end{figure}
\newpage

\noindent
 from the recent CHOOZ 
\cite{CHOOZ} and 
Super--K results \cite{Superk}), confirm or exclude
Majorana neutrinos as hot dark matter in the universe as well as to 
test SUSY models, leptoquarks and
right--handed W--masses comparable to the LHC \cite{Kla98,gen}. The required
purity levels for the liquid nitrogen are (except for $^{222}$Rn and 
$^{40}$K) less
stringent than already obtained by the Borexino Collaboration for the liquid 
scintillator \cite{borex}. 
A ten ton version would probe neutrino masses even down to $10^{-3}$ eV. 
This way it tests the large angle MSW solution -- and in less hierarchical
scenarios even the small angle solution -- of the solar neutrino 
problem for all oscillation channels (see Fig. 5). 
As direct dark matter detection experiment it would allow to test almost 
the entire MSSM parameter space (see Fig. 4) already in a first step using 
only 100 kg of enriched or even natural Ge \cite{Kla98,gen}. 

\section{Conclusions}
Neutrinoless double beta decay and dark matter search belong to the most 
sensitive approaches with great perspectives 
to test particle physics beyond the SM.
 
The possibilities to use $0\nu\beta\beta$ decay for constraining 
neutrino masses,
left--right--symmetric models, SUSY and leptoquark scenarios, as well as
effective lepton number violating couplings, 
have been reviewed. Experimental 
limits on $0\nu\beta\beta$ decay are not only complementary to 
accelerator experiments but at least in some cases competitive or superior
to the best existing direct search limits. 
The Heidelberg--Moscow experiment
has reached the leading position among double beta decay experiments and as
the first of them yields results in the sub--eV range for the neutrino 
mass. 

Direct WIMP detection experiments 
can compete with recent and future accelerator 
experiments in the search for SUSY and experiments using double beta 
technology belong to the most promising approaches in this field of research.
Here the Heidelberg--Moscow experiment yields the most stringent limits 
on spin--independent WIMP interactions using only raw data.

A further large breakthrough, 
both for double beta decay and dark matter search, 
will be possible realizing the GENIUS proposal.

\end{document}